\documentclass[12pt]{article}
\usepackage{amsthm,amssymb,amsmath}
\usepackage{graphicx}
\newcommand{\snu}{\tilde N}
\newcommand{\sH}{\tilde H}
\newcommand{\AddrUAM}{
  \it \small Departamento de F\'{\i}sica Te\'orica C-XI {\rm and}  Instituto de F\'{\i}sica Te\'orica UAM-CSIS
}
\newcommand{\AddrIFT}{
  \it \small Universidad
    Aut\'onoma de Madrid, Cantoblanco, E-28049 Madrid, Spain
  }
\begin{document}
\begin{titlepage}
\begin{center}
\textbf{\LARGE Inverse decays and the relic density of the sterile sneutrino}

\vspace{1cm}
Carlos E. Yaguna\\[3mm]
\AddrUAM \\
\AddrIFT\\

\end{center}
\begin{abstract}
We consider a weak scale supersymmetric seesaw model  where the Higgsino is the next-to-lightest supersymmetric particle and the right-handed sneutrino is the dark matter candidate. It is  shown that, in this model, inverse decays, which had been previously neglected, may suppress the sneutrino relic density by several orders of magnitude. After including such processes and numerically solving the appropriate Boltzmann equation, we study the dependence of the relic density on the $\mu$ parameter, the sneutrino mass, and the neutrino Yukawa coupling. We find that, even though much smaller than in earlier calculations, the sneutrino relic density is still larger than the observed dark matter density. 
\end{abstract}
\end{titlepage}
\section{Introduction}
The supersymmetric version of the seesaw mechanism is an attractive candidate for physics beyond the Standard Model. On the one hand, it includes the seesaw mechanism, which postulates the  existence of right-handed neutrino fields and has become the most popular framework to account for neutrino masses. The seesaw is able to accommodate the experimental data on neutrino masses and mixing \cite{Yao:2006px}, explaining naturally the small neutrino mass scale.  On the other hand, it embraces low energy supersymmetry, with its rich phenomenology and its well known virtues. In fact, the minimal supersymmetric Standard Model solves the hierarchy problem, achieves the unification of the gauge couplings, and contains a dark matter candidate: the lightest supersymmetric particle.

The lightest sneutrino is a new dark matter candidate present in the supersymmetric seesaw. Being a mixture of left-handed and right-handed sneutrino fields, the lightest sneutrino will have different properties depending on its composition in terms of interactions eigenstates. In general,  three different kind of sneutrinos can be envisioned: a dominantly left-handed one, a mixed sneutrino, or a dominantly right-handed one. A dominantly left-handed sneutrino is not a good dark matter candidate. They are ruled out by experimental searches \cite{Ahlen:1987mn} and tend to have a too small relic density \cite{Falk:1994es}. A mixed sneutrino can be compatible with the observed dark matter density as well as with present bounds from direct searches \cite{ArkaniHamed:2000bq,Arina:2007tm}. The required mixing is obtained at the expense of  a  large  neutrino trilinear coupling, which is not allowed in typical models of supersymmetry breaking. A dominantly right-handed sneutrino is the final possibility, the one we will be concerned with throughout this paper. A right-handed sneutrino, being essentially sterile,  interacts with other particles mainly through the neutrino Yukawa coupling. Could  such a sterile sneutrino account for the observed dark matter density?

Gopalakrishna, Gouvea, and Porod, in \cite{Gopalakrishna:2006kr}, studied that possibility within the same scenario we are considering here. They showed that self-annihilations of right-handed sneutrinos as well as co-annihilations with other particles are too weak to keep the sneutrinos in equilibrium with the thermal plasma in the early Universe. They also found that the production of sneutrinos in the decay of other supersymmetric particles gives a too large contribution to the relic density. They concluded, therefore, that in the standard cosmological model right-handed sneutrinos cannot explain the dark matter of the Universe.

Even though generally valid, that conclusion is not guaranteed if the mass difference between the Higgsino and the sneutrino is small. In that case,  inverse decays, such as $\snu+L\to \sH$, contribute to the annihilation of sneutrinos and therefore to the reduction of the sneutrino relic density. Such possibility was not taken into account in \cite{Gopalakrishna:2006kr}. In this paper, we will focus on models with a Higgsino NLSP and show that  inverse processes cannot be neglected, for they suppress the sneutrino relic density by several orders of magnitude. Then, we will  reexamine whether the sterile sneutrino can explain the dark matter of the Universe in the standard cosmological model.

In the next section we briefly review the supersymmetric seesaw model and  show that sterile sneutrinos arise naturally in common scenarios of supersymmetry breaking. Then, in section \ref{sec:3}, we will include inverse decays into the  Boltzmann equation that  determines the sneutrino abundance. It is then shown that inverse decays are indeed relevant; they cause a significant reduction of the relic density. In section \ref{sec:4}, we study the relic density as a function of the neutrino Yukawa coupling, the sneutrino mass, and the Higgsino-sneutrino mass difference. There, we will obtain our main result: the suppression effect of inverse decays, though  important, is not enough to bring the sneutrino relic density down within the observed range. In the final section we will review our study and present our conclusions. 

\section{The model}\label{sec:2}
We work within the supersymmetric version of the seesaw mechanism, where the field content of the MSSM is supplemented with a right-handed neutrino superfield $N$ per generation. The superpotential then reads
\begin{equation} 
W=W_{MSSM}+\frac12M_N^{IJ}N^IN^J+Y_\nu^{IJ} H_uL^IN^J
\label{superp}
\end{equation}
where, as usual, we have assumed R-parity conservation and renormalizability. $M_N$ is the Majorana mass matrix of right-handed neutrinos and $Y_\nu$ is the matrix of neutrino Yukawa couplings. Without loss of generality $M_N$ can be chosen to be real and diagonal. $Y_\nu$  is in general complex but we will assume, for simplicity,  that it is real. $M_N$ and $Y_\nu$ are new free parameters of the model; they are to be determined or constrained from experimental data.

After electroweak symmetry breaking, the above superpotential  generates  the following neutrino mass terms
\begin{equation}
\mathcal{L}_{\nu\,mass}=-v_uY_\nu \nu N-\frac12M_NNN+h.c.
\end{equation}
If $M_N\gg v_u Y_\nu$, the light neutrino mass matrix, $m_\nu$, is then given by the seesaw formula
\begin{equation}
m_\nu=-m_DM_N^{-1}m_D^T,
\label{seesaw}
\end{equation}
with $m_D=v_uY_\nu$ being the Dirac mass. Since $m_\nu$ is partially known from neutrino oscillation data,  equation (\ref{seesaw}) is actually a  constraint on the possible values of $Y_\nu$ and $M_N$. It is a weak constraint though; and  it allows $M_N$ to vary over many different scales. In this paper we consider what is usually known as a seesaw mechanism at the electroweak scale. That is, we assume that $M_N\sim 100$ GeV. Thus, since the neutrino mass scale is around $m_\nu\sim 0.1$ eV, the typical neutrino Yukawa coupling is 
\begin{equation}
Y_\nu\sim 10^{-6}\,,
\end{equation}  
or around the same order of magnitude as the electron Yukawa coupling. Notice that this value of $Y_\nu$ is a consequence of the seesaw mechanism at the electroweak scale. In other frameworks, such as Dirac neutrinos or seesaw at much higher energies, $Y_\nu$ takes different values. We will not consider such possibilities here.

The new soft-breaking terms of the supersymmetric seesaw model are given by
\begin{equation}
\mathcal{L}_{soft}=-(m_N^2)^{IJ}\tilde N_R^{*I}\tilde N_R^J+\left[(m_B^2)^{IJ}\tilde N_R^I\tilde N_R^J-A_\nu^{IJ}h_u\tilde L^I\tilde N_R^J+h.c.\right]\,.
\label{lsoft}
\end{equation}
They include sneutrino mass terms as well a trilinear interaction term. For simplicity, we will assume that $m_N^2$, $m_B^2$, and $A_\nu$ are real.

To  study the sneutrino mass terms resulting from (\ref{superp}) and (\ref{lsoft}) it is convenient to suppress the generation structure; that is, to work with one fermion generation only. It is also useful to introduce the real fields $\tilde\nu_1$, $\tilde\nu_2$, $\tilde N_1$ and $\tilde N_2$ according to the relations
\begin{eqnarray}
\tilde\nu_L=\frac{1}{\sqrt2}\left(\tilde \nu_1+i\tilde \nu_2\right)\,,
\tilde N_R=\frac{1}{\sqrt2}\left(\tilde N_1+ i \tilde N_2\right).
\end{eqnarray}
Indeed, in the basis $(\tilde \nu_1,\tilde N_1,\tilde \nu_2,\tilde N_2)$ the sneutrino mass matrix takes a block diagonal form
\begin{equation}
\mathcal{M}_{\tilde\nu}=\left(\begin{array}{cccc} m_{LL}^2 & m_{RL}^{2}+m_DM_N & 0 &0\\
m_{RL}^2+m_D M & m_{RR}^2-m_B^2 & 0 &0 \\
0 & 0& m_{LL}^2 & m_{RL}^{2}-m_DM_N\\
0 & 0& m_{RL}^2-m_DM_N & m_{RR}^2+m_B^2\end{array}\right)
\label{eq:mv}
\end{equation}
where $m_{LL}=m_{\tilde L}^2+m_D^2+0.5m_Z^2\cos2\beta $, $m_{RR}^2=M_N^2+m_N^2+m_D^2$, and $m_{LR}^2=-\mu v_dY_N+v_uA_\nu$.
This matrix can be diagonalized by  a unitary rotation with a mixing angle  given by
\begin{equation}
\tan 2\theta_{1,2}^{\tilde \nu}=\frac{2(m_{RL}^2\pm m_DM)}{m_{LL}^2-(m_{RR}^2\mp m_B^2)},
\label{eq:mix}
\end{equation}
where the top sign corresponds to $\theta_1$ --to the mixing between $\tilde \nu_1$ and $\tilde N_1$-- whereas the bottom sign corresponds to $\theta_2$.

Since $\mathcal{M}_{\tilde\nu}$ is independent of gaugino masses, there is a region in the supersymmetric parameter space where the lightest sneutrino, obtained from  (\ref{eq:mv}), is the lightest supersymmetric particle (LSP) and consequently the dark matter candidate. That is the only region we will consider in this paper.

The lightest sneutrino is a mixture of left-handed and right-handed sneutrino fields. Depending on its gauge composition, three kinds of sneutrinos can be distinguished: a dominantly left-handed sneutrino, a mixed sneutrino, and a dominantly right-handed sneutrino. A dominantly left-handed sneutrino is not a good dark matter candidate for it is already ruled out by direct dark matter searches. These sneutrinos also have  large interactions cross sections and tend to annihilate efficiently in the early universe, typically yielding a too small relic density. A mixed sneutrino may be a good dark matter candidate. By adjusting the sneutrino mixing angle, one can simultaneously suppress its annihilation cross section, so as to obtain the right relic density, and the sneutrino-nucleon cross section, so as to evade present constraints from direct searches. A detailed study of models with  mixed sneutrino dark matter was presented recently in \cite{Arina:2007tm}. A major drawback of these models is that the required mixing may be incompatible with certain scenarios of supersymmetry breaking, such as gravity mediation. The third possibility, the one we consider, is a lightest sneutrino which is predominantly right-handed. That is, a \emph{sterile sneutrino}.

A sterile sneutrino is actually unavoidable in supersymmetry breaking scenarios where the trilinear couplings are proportional to  the corresponding Yukawa matrices, such as the constrained Minimal Supersymmetric Standard Model (CMSSM)\cite{Yao:2006px}.  In these models
\begin{equation}
A_\nu=a_\nu Y_\nu m_{soft}
\end{equation} 
where $m_{soft}\sim 100$ GeV is a typical supersymmetry breaking mass and $a_\nu$ is an order one parameter. Because $Y_\nu$ is small, $A_\nu$ is much smaller than the electroweak scale,
\begin{equation}
 A_\nu\sim 100 \mathrm{keV}\,.
\end{equation}
Hence, from equation (\ref{eq:mix}), the mixing angle between $\tilde \nu_i$ and $\tilde N_i$ is also very small
\begin{equation}
 \sin\theta_i\sim 10^{-6}\,.
\end{equation}
Thus, we see how in these models the small $Y_\nu$ translates into a small trilinear coupling $A_\nu$ that in turn leads to a small mixing angle --to a sterile sneutrino. Sterile sneutrinos are also expected in other supersymmetry breaking mechanisms that yield a  small $A_\nu$ at the electroweak scale.  

Since the mixing angle is small, we can extract the sterile neutrino mass directly from (\ref{eq:mv}). It is given by
\begin{equation}
m_{\snu}^2=m_{RR}^2-m_{B}^2\approx M_N^2+m_N^2-m_B^2
\end{equation} 
where we have neglected the Dirac mass term in the last expression. $m_{\snu}$ is thus expected to be at the electroweak scale. In the following, we will consider $m_{\snu}=m_{LSP}$ as a free parameter of the model.   

To summarize, the models we study consist of the MSSM plus an electroweak scale seesaw mechanism that accounts for neutrino masses. Such models include a new dark matter candidate: the lightest sneutrino. In common scenarios of supersymmetry breaking, the lightest sneutrino, which we assume to be the dark matter candidate, turns out to be a dominantly right handed sneutrino, or a sterile sneutrino.  In the following, we will examine whether such a \emph{sterile} sneutrino may account for the dark matter of the Universe.

\section{The $\snu$ relic density}\label{sec:3}
To determine whether the sterile sneutrino can explain the dark matter of the universe we must compute its relic density $\Omega_{\snu}h^2$ and compare it with the observed value $\Omega_{DM}h^2=0.11\cite{Dunkley:2008ie}$. This question  was already addressed in \cite{Gopalakrishna:2006kr}. They showed that, due to their weak interactions, sneutrinos are unable to reach thermal equilibrium in the early Universe. In fact, both the self-annihilation and the co-annihilation cross section are very suppressed. They also noticed that sneutrinos could be produced in the decays of other supersymmetric particles and found that such decay contributions  lead to a relic density  several  orders of magnitude larger than observed. Thus, they concluded, sterile sneutrinos can only be non-thermal dark matter candidates.

That conclusion was drawn, however, without taking into account inverse decay processes. We now show that if the Higgsino-sneutrino mass difference is small\footnote{If it is large the results in \cite{Gopalakrishna:2006kr} would follow.},  inverse decays may suppress the sneutrino relic density  by several orders of magnitude.  To isolate this effect, only models with a Higgsino NLSP are considered in the following. We then reexamine the possibility of having a sterile sneutrino as a thermal dark matter candidate within the standard cosmological model.

In the early Universe, sterile sneutrinos are mainly created through the decay $\sH\to \snu+L$, where $\sH$ is the Higgsino and $L$ is the lepton doublet. Alternatively, using the mass-eigenstate language, one may say that sneutrinos are created in the decay of neutralinos ($\chi^0\to \snu +\nu$) and charginos ($\chi^\pm\to \ell^\pm +\snu$). These decays are all controlled by the neutrino Yukawa coupling $Y_\nu$. Other  decays, such as $\tilde\ell\to \snu f f'$ via $W^\pm$, also occur but the Higgsino channel dominates. Regarding annihilation processes, the most important one is the inverse decay  $\snu+L\to\sH$. In fact, the sneutrino-sneutrino annihilation cross section is so small that such process never reaches equilibrium. And a similar result holds for the sneutrino coannihilation cross section. We can therefore safely neglect annihilations and coannihilations in the following. Only decays and inverse decays contribute to the sneutrino relic density.

The Boltzmann equation for the sneutrino distribution function $f_{\snu}$ then reads:
\begin{align}
\label{boltzmann}
\frac{\partial f_{\snu}}{\partial t}-H\frac{|\mathbf{p}|^2}{E}\frac{\partial f_{\snu}}{\partial E}=\frac{1}{2 E_{\snu}}\int & \frac{d^3p_L}{(2\pi)^3 2E_L}\frac{d^3p_{\sH}}{(2\pi)^3 2E_{\sH}}|\mathcal{M}_{\tilde H\to L\snu}|^2\\ \nonumber
 & (2\pi)^4 \delta^4(p_{\tilde H}-p_L-p_{\snu})\left[f_{\sH}-f_L f_{\snu}\right]
\end{align} 
where $H$ is the Hubble parameter and $f_{\sH}$, $f_{L}$ respectively denote the $\sH$ and $L$ distribution functions. Other dark matter candidates, including  the neutralino, have large elastic scatterings cross sections with the thermal plasma that  keep them  in \emph{kinetic} equilibrium during the freeze out process. Their distribution functions are then proportional to those in \emph{chemical} equilibrium and the Boltzmann equation can be written as an equation for the number density instead of the distribution function \cite{Gondolo:1990dk}. For sterile sneutrinos, on the contrary, the elastic scattering is a slow process --being suppressed by the Yukawa coupling-- and  kinetic equilibrium is not guaranteed. Hence, we cannot write (\ref{boltzmann}) as an equation for the sneutrino number density $n_{\snu}$ and must instead solve it for $f_{\snu}$. 
  
If the condition $f_{\snu}\ll 1$ were satisfied, inverse processes could be neglected and a simple equation relating the sneutrino number density to the Higgsino number density could be obtained. That is the case, for instance, in supersymmetric scenarios with Dirac mass terms only \cite{Asaka:2005cn}. In such models, the neutrino Yukawa coupling is very small, $Y_\nu\sim 10^{-13}$, and sneutrinos never reach chemical equilibrium. But for the range of parameters we consider, $Y_\nu\sim 10^{-6}$, the condition $f_{\snu}\ll 1$ is not satisfied.

Since equation (\ref{boltzmann}) depends also on the Higgsino distribution function, one may think that  it is necessary to write the Boltzmann equation for  $f_{\sH}$  and then solve the resulting system  for $f_{\snu}$ and $f_{\sH}$. Not so. Higgsinos, due to their gauge interactions, are kept in thermal equilibrium --by self-annihilation processes-- until low temperatures, when they decay into $\snu+L$ through the $Y_\nu$ suppressed interaction. It is thus useful to define a \emph{freeze-out} temperature, $T_{f.o.}$, as the temperature at which these two reaction rates become equal. That is, 
\begin{equation}
n_{\sH}\langle\sigma_{\sH\sH}v\rangle|_{T_{f.o.}}=\Gamma(\sH\to \snu +L)|_{T_{f.o.}}\,,
\label{eq:fo}
\end{equation}
where $n_{\sH}$ is the Higgsino number density and $\langle\sigma_{\sH\sH}v\rangle$ is the thermal average of the Higgsino-Higgsino annihilation rate into light particles. $T_{f.o.}$ marks the boundary between two different regimes. For $T>T_{f.o.}$ Higgsinos are in equilibrium and annihilate efficiently. The Higgsinos produced in the inverse decays, in particular,  easily annihilate with thermal Higgsinos into light particles. The inverse process is thus effective. In contrast,    for $T<T_{f.o.}$ Higgsinos mostly decay into the LSP and inverse decays cannot deplete the sneutrino abundance. The final state Higgsinos simply decay back into sneutrinos: $\snu+L\to\sH\to \snu+L$.
Below $T_{f.o.}$, therefore, the total number of sneutrinos plus Higgsinos remains constant.  Thus, we only need to integrate equation (\ref{boltzmann2}) until $T_{f.o.}$, a  region in which Higgsinos are in equilibrium.
 
Assuming a Maxwell-Boltzmann distribution, $f(E)\propto \exp(-E/T)$, for Higgsinos and leptons and neglecting lepton masses, the integrals in (\ref{boltzmann}) can be evaluated analytically to find
\begin{equation}
\frac{\partial f_{\tilde \nu}}{\partial t}-H\frac{|\mathbf{p}|^2}{E}\frac{\partial f_{\tilde \nu}}{\partial E}=\frac{|\mathcal{M}_{\tilde H\to L\snu}|^2 T}{16\pi E_{\snu}|\mathbf{p}_{\snu}|}\left(e^{-E_{\snu}/T} -f_{\snu}\right) \left[e^{-E_{-}/T}-e^{-E_{+}/T}\right]
\label{boltzmann2}
\end{equation} 
where
\begin{align}
E_\pm&=\frac{m_{\sH}^2-m_{\snu}^2}{2m_{\snu}^2}(E_{\snu}\pm|\mathbf{p}_{\snu}|).
\end{align}

In the following we will solve equation (\ref{boltzmann2}) to obtain the sneutrino abundance, $Y_{\snu}=n_{\snu}/s$, and the sneutrino relic density, $\Omega_{\snu}h^2$.  The sneutrino abundance today  will be given by 
\begin{equation}
Y_{\snu}|_{T_0}=Y_{\snu}|_{T_{f.o.}}+Y_{\sH}|_{T_{f.o.}},
\end{equation}
where the second term takes into account that the Higgsinos present at freeze-out will decay into sneutrinos. The sneutrino relic density today is then obtained as 
\begin{equation}
\Omega_{\snu}h^2=2.8\times10^{10} Y_{\snu} \frac{m_{\snu}}{100\mathrm{GeV}}.
\end{equation}
 The only parameters that enter directly in the computation of the sneutrino relic density are the Yukawa coupling, the sneutrino mass, and the Higgsino mass, which we take to be given by the $\mu$ parameter --$m_{\sH}=\mu$. All other supersymmetric particles besides $\snu$ and $\sH$ are assumed to be heavier, with $m_{susy}\sim 1$ TeV. To determine the freeze-out temperature, equation (\ref{eq:fo}), we also need to know the Higgsino annihilation rate into Standard Model particles. We use the DarkSUSY package \cite{Gondolo:2004sc} to extract that value. Regarding the initial conditions, we  assume that at high temperatures ($T\gg m_{\sH}$) the sneutrino distribution function is negligible $f_{\snu}\sim 0$. Finally, we assume that the early Universe is described by the standard cosmological model.
\begin{figure}[t]
\begin{center}
\includegraphics[scale=0.4,angle=-90]{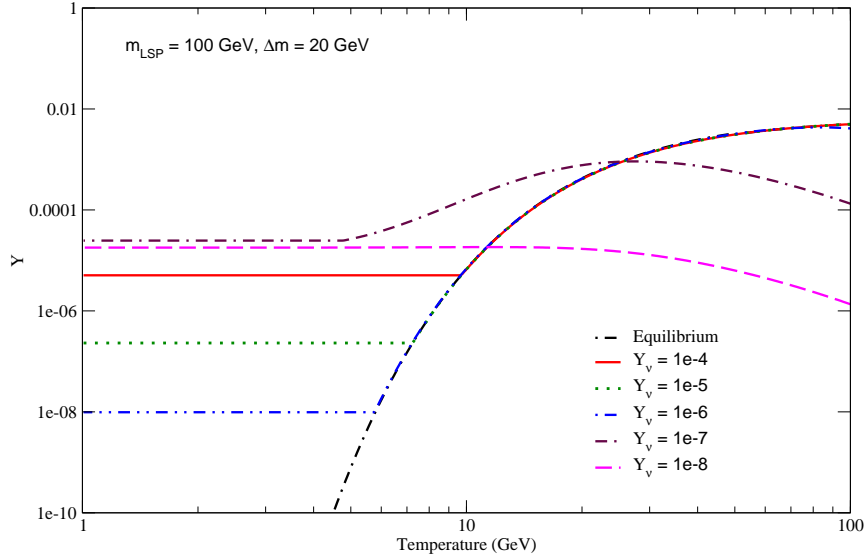}
\caption{\small The role of the neutrino Yukawa coupling on the sterile sneutrino abundance. The figure shows $Y$ as a function of the temperature for different values of $Y_\nu$. The sneutrino mass is $100$GeV while $\mu=120$GeV.}
\label{figure1}
\end{center}
\end{figure}

Once decays and inverse decays are included in the $\snu$ Boltzmann equation, two questions naturally come to mind. First, for what values of $Y_\nu$ are inverse decays relevant? Second, can decays and inverse decays bring the sneutrinos into equilibrium? To answer these questions we show in figure \ref{figure1} the sneutrino abundance as a function of the temperature for $m_{\snu}=100$ GeV, $m_{\sH}=120$ GeV, and different values of $Y_\nu$. Notice that for $Y_\nu=10^{-8}$ inverse processes are negligible and the sneutrino abundance simply grows with  temperature. In that region, for $Y_\nu\lesssim 10^{-8}$, the sneutrino relic density is proportional to $Y_\nu^2$. From the figure we see that for $Y_\nu=10^{-7}$ the inverse process leads to a reduction of the sneutrino abundance around $T=20$ GeV. The Yukawa interaction is not yet strong enough to bring the sneutrinos into equilibrium. For $Y_\nu=10^{-6}$ sneutrinos do reach equilibrium  and then decouple at lower temperatures. For even larger Yukawa couplings, $Y_\nu=10^{-5},10^{-4}$, equilibrium is also reached but the decoupling occurs at higher temperatures. In that region, the relic density also increases with the Yukawas. Thus, for $Y_\nu\sim 10^{-6}$ inverse decays not only are relevant, they are strong enough to thermalize the sneutrinos.

\begin{figure}[t]
\begin{center}
\includegraphics[scale=0.4,angle=-90]{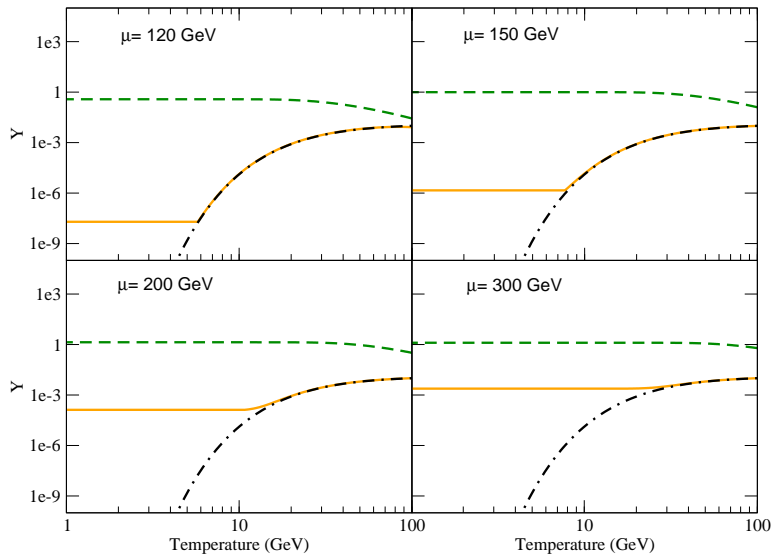}
\end{center}
\caption{\small The effect of the inverse process on the sneutrino relic density. The panels show the resulting sneutrino abundance $Y=n/s$ as a function of the temperature for $m_{\snu}=100$GeV and different values of $\mu$. The full line is the result obtained including the inverse process whereas  the dashed line is the result without including them. The dash-dotted line shows the sneutrino equilibrium abundance.}
\label{figure2}
\end{figure}

Figure \ref{figure2} directly compares  the resulting sneutrino abundance with and without including the inverse process. The full line corresponds to the correct result, taking into account the direct and the inverse process. The dashed line, instead, shows the result for the direct process only, that is the sneutrino abundance according to \cite{Gopalakrishna:2006kr}. The sneutrino mass was taken to be $100$GeV and $Y_\nu$ was set to $10^{-6}$. The Higgsino mass is different in each panel and includes values leading to strong and  mild degeneracy as well as no-degeneracy at all between the sneutrino and the Higgsino. Notice that the correct final abundance, and consequently the resulting relic density, is always several orders of magnitude below the value predicted in \cite{Gopalakrishna:2006kr}. Even for the case of a large mass difference, we find a suppression of 3 orders of magnitude in the relic density. And as the mass difference shrinks the suppression becomes larger, reaching about $6$ orders of magnitude for $\mu=150$ and about $7$ orders of magnitude for $\mu=120$GeV. We thus see that over the whole parameter space the inverse process has a  large suppression effect on the sneutrino relic density.

\section{Results}
\label{sec:4}

\begin{figure}[t]
\begin{center}
\includegraphics[scale=0.4,angle=-90]{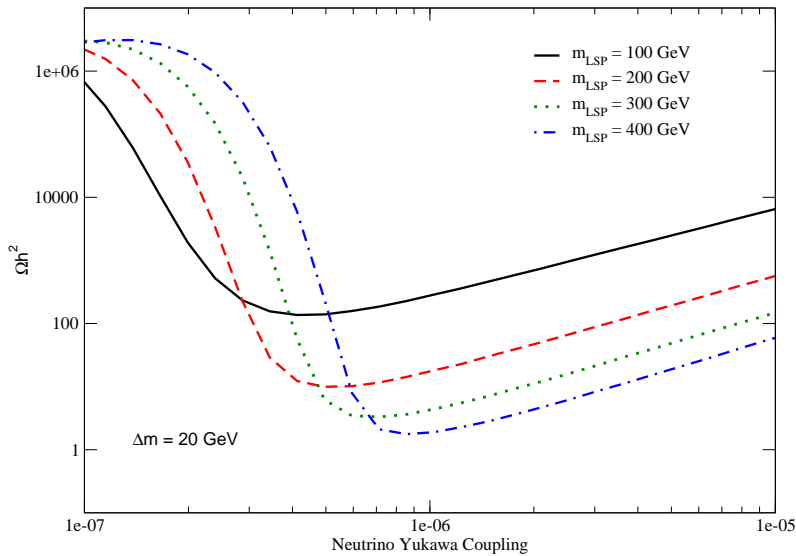}
\caption{\small The sneutrino relic density as a function of the neutrino Yukawa coupling for different values of $m_{\snu}$ and $\Delta m=20$GeV.}
\label{figure3}
\end{center}
\end{figure}

So far we have found that the inverse decay process $\snu+L\to\sH$ leads to a suppression  of the sneutrino  relic density. It remains to be seen  whether such suppression is strong enough to bring the relic density down to the observed value. That is, we will now study the dependence of the relic density with the sneutrino mass, the Higgsino-sneutrino mass difference, and the neutrino Yukawa coupling to find the region of the parameter space that satisfies the condition $\Omega_{\snu}h^2=\Omega_{DM}h^2$.

Figure \ref{figure3}  shows the sneutrino relic density as a function of the neutrino Yukawa coupling  and different values of the sneutrino mass. The Higgsino-sneutrino mass difference ($\Delta m=m_{\sH}-m_{\snu}$) was set to $20$ GeV. Larger values would only increase the relic density --see figure \ref{figure2}. Notice that, for a given sneutrino mass,  the relic density initially decreases rather steeply reaching a minimum value at $Y_\nu\lesssim 10^{-6}$ and then increases again. From the figure we also observe that the smallest value of the relic density is obtained for $m_{\sH}=400$ GeV, that is, when the percentage mass difference is smaller. In any case, the relic density is always larger than $1$, too large to be compatible with the observations.

\begin{figure}[t]
\begin{center}
\includegraphics[scale=0.4,angle=-90]{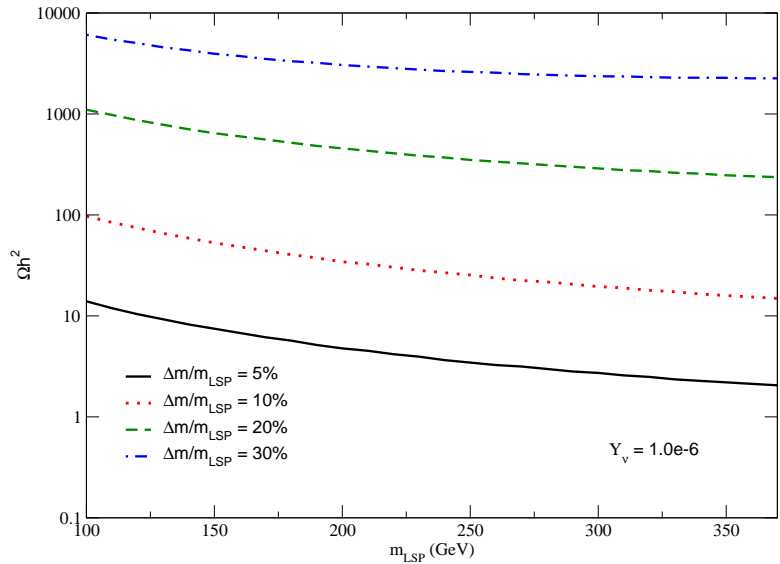}
\caption{\small The sneutrino relic density as a function of the sneutrino mass for $Y_\nu=10^{-6}$ and different values of $\Delta m/m_{\snu}$. As expected the smaller the mass difference the smaller the relic density is.}
\end{center}
\label{fig:msn}
\end{figure}

This result is confirmed in figure \ref{fig:msn} when we display the relic density as a function of the sneutrino mass for $Y_{\snu}=10^{-6}$ and different values of $\Delta m/m$. In agreement with the previous figure, we see that the smaller the percentage mass difference, the smaller the relic density is. Yet, $\Omega_{\snu}h^2$ is always larger than $1$. We have verified that this conclusion is robust. Neither larger sneutrino masses nor different Yukawa couplings lead to the correct value of the relic density.

\section{Conclusions}
We studied the possibility of explaining the dark matter with a sterile sneutrino in a supersymmetric model consisting of the MSSM supplemented with a seesaw mechanism at the weak scale. We showed that if the Higgsino is the NLSP inverse decays play a crucial role in the computation of the sneutrino relic density, suppressing it for several orders of magnitude. We wrote down and numerically solved the correct Boltzmann equation that determines the sneutrino abundance and studied the resulting relic density as a function of the sneutrino mass, the neutrino Yukawa coupling and the Higgsino-sneutrino mass difference. We found that the sterile sneutrino relic density, even though much smaller than previously believed, is still larger than the observed dark matter density. In this scenario, therefore, the sterile sneutrino is not a thermal  dark matter candidate.

\section*{Acknowledgments}
I am  supported by the \emph{Juan de la Cierva} program of the Ministerio de Educacion y Ciencia of Spain, by Proyecto Nacional FPA2006-01105, and by the Comunidad de Madrid under Proyecto HEPHACOS S-0505/ESP-0346. I would like to thank W. Porod and Ki-Young Choi for comments and suggestions.

\end{document}